\documentclass[runningheads]{llncs}
\usepackage{amsmath,amssymb,amsfonts}
\usepackage{algorithmic}
\usepackage{graphicx}
\usepackage{graphicx}
\usepackage{subcaption}
\usepackage[sorting=none]{biblatex}
\usepackage{ntheorem}
\theoremseparator{:}
\newtheorem{hyp}{Null Hypothesis}

\begin{document}
\title{A Study on Workload Assessment and Usability of Wind-Aware User Interface for Small Unmanned Aircraft System Remote Operations\thanks{The work is supported by the U.S. National Science Foundation (NSF) under award number 1925147.}}
\titlerunning{Accepted at HCII 2023: Late breaking papers}
\author{Asma Tabassum \and
He Bai \and
Nicoletta Fala}
\authorrunning{A. Tabassum et al.}
%
\institute{Oklahoma State University, Stillwater Ok 74075, USA 
\email{\{asma.tabassum,he.bai,nfala\}@okstate.edu}}
\maketitle              
\begin{abstract}
This study evaluates pilots' cognitive workload and situational awareness during remote small unmanned aircraft system operations in different wind conditions.  To complement the urban air mobility concept that envisions safe, sustainable, and accessible air transportation, we conduct multiple experiments in  a realistic wind-aware simulator-user interface pipeline. Experiments are performed with basic and wind-aware displays in several wind conditions to assess how complex wind fields impact pilots' cognitive resources.  Post-hoc analysis reveals that providing pilots with real-time wind information improves situational awareness while decreasing cognitive workload.

\end{abstract}
\section{Introduction}
The idea of urban air mobility (UAM) encompasses air transportation within and above a city and a subset of advanced air mobility~\cite{reiche2021initial}, which aspires to produce a safe, secure, and efficient air traffic operation. With the aggressive integration of unmanned aircraft system (UAS) into the National Airspace System, more than 250 prototypes of vertical take-off and landing (VTOL), electric, and autonomous aircraft are being designed and tested~\cite{reiche2021initial}. Even with all the growing attention and a global expected growth of $13.8\%$ by 2025~\cite{rebensky2022impact}, the designs for UAS and large-scale integration are being challenged by environmental uncertainties~\cite{UPS}. One of these critical environmental hurdles is turbulent wind, especially in urban settings. Around 52 percent of respondents (out of 1702 people) express an increased level of fear and concern while flying in the turbulent wind in UAM~\cite{reiche2021initial}. For aerospace applications, control design for a standalone airborne system itself demands more scrutiny and requires guaranteed operability in a dynamic environment. This becomes more challenging for UAS remote control operations, where the pilot cannot analyze the dynamic environment onboard. Due to sensory isolation caused by the Ground Control Station (GCS) being located on the ground, the UAS pilot is deprived of possible vestibular and onboard visual senses. However, the UAS pilot must acquire and interpret the equivalent level of awareness and information (as of a crewed aircraft) through sensors and interfaces regardless of autonomous or manual operation. On top of that, challenges imposed by uncertainties degrade UAS operation and navigation tasks. Therefore, compensating for the reduced situational awareness (SA) is a major challenge in UAS GCS User Interface (UI) design. Situational awareness refers to the operator's internal model of the surrounding world around them at any time. Addressing the drawbacks of reduced SA in UAS is not straightforward as it relates to including additional information, layouts, and audio-visual inputs into the UI. As a consequence, UAS UI design incurs major pitfalls such as \begin{itemize}
    \item misidentification of operational information during the design phase~\cite{vincenzi2015unmanned},
    \item addition of irrelevant information leading to additional cognitive processing~\cite{tvaryanas2008recurrent}
    \item inaccurate representation of information leading to inadequate/incorrect responses of the operator~\cite{hobbs2010unmanned}.
\end{itemize}
Thus, poor display design and poor information presentation increase task overhead and significantly impact mission quality and operator performance.

Modern UI designs have focused on user-centric designs~\cite{chammas2015closer} which emphasize a user's needs and application requirements. UAS UI design through research, simulation, and usability testing could potentially satisfy users' needs and design standards~\cite{haritos2017study}. While the research effort in the interaction between human pilot and autopilot in crewed aircraft is more mature, less attention has been expended to investigate interaction strategies associated with remote UAS pilot and onboard command and control~\cite{terwilliger2014advancement}. An initial design guide for interface design that involves viewpoint design, control level design, and autonomy level design is illustrated by FAA~\cite{williams2007assessment}. In~\cite{maybury2012usable}, the author recommends an iterative task analysis throughout the design process to better understand key task components, user needs, and the user's mental representation of the displayed information. The necessity of the assessment of the pilot's cognitive states with autonomy is reported in~\cite{jimenez2016user}. Other studies~\cite{vinot2016could,monk2015effects} also discuss the design criteria for the sUAS interface. In our previous work~\cite{tabassum2022preliminary}, we adopt a user-centered design methodology to develop a wind-aware UI and simulation pipeline for small UAS (sUAS). We identify information components and derive specific display designs based on a focus group study with subject matter experts (SME). Our subject matter experts are four pilots, and three of them have FAR 107 certifications. One of the biggest concerns of SMEs is the lack of wind velocity information in the current off-the-shelf displays. They also mention that operators mostly rely on local wind predictions and do not have access to wind information through the GCS UI during flight. Based on experts' suggestions, we implement a design that overlays wind information into the QGroundcontrol UI to accommodate a wind-aware framework throughout small UAS missions.

Literature indicates that about $69\%$ of the UAS mishaps (damage or loss of platform) are caused by human factors~\cite{waraich2013minimizing}. In this work, we aim to address the rising operational and navigational challenges that turbulent wind imposes on human pilots and investigate how wind-aware UI may reduce a
remote pilot’s subjective cognitive load and improve situational awareness. We expect that given that a human pilot has adequate knowledge (training) of the system and tasks, including real-time information on wind will assist  accurate perception of the vehicle and environment, reduce cognitive load, and enhance the piloting interaction experience. Toward this end, we evaluate the usability of wind-aware UI and assess the cognitive workload of the participants while flying with the UI.  Usability refers to the quality of a UI that is easy to learn, has effective use, and is enjoyable from the user’s perspective~\cite{preece2015interaction}.

The main goal of our design is to improve situational awareness through the easy and effective use of wind-aware UI. We test usability by evaluating the improvement in situational awareness while reducing cognitive workload. We use a fully manual mode for the experiment. This mode does not support stable hover or station keeping in the presence of disturbance, i.e., the pilot must keep it stable using manual input. Manual modes are specifically used during search and rescue operations or to capture pictures of particular objects~\cite{rebensky2022impact}. While relatively stable manual control or GPS-guided control options are available in off-the-shelf platforms, the rationale behind choosing this mode is twofold. First, this mode allows us to let the pilot experience the wind as is so that we can capture relatively accurate cognitive exhaustion that the wind imposes on the pilot. We use pilots' cognitive data in our subjective analysis. Second, this mode is expected to apprehend how individual flight experiences cause different performances. These effects are expected to be captured by flight record data, e.g., states and input, and will be assessed for objective analysis in subsequent studies. Based on the subjective interpretation of participants, this study aims to answer the following three questions:
\begin{itemize}
    \item Does wind significantly increase the workload for remote pilots?
    \item Does wind-aware UI alleviate pilots' cognitive loading?
    \item Does wind-aware UI improve subjective situational awareness?    
\end{itemize}
Based on our research questions, we deduce three hypotheses and test them with the experimental data.  The rest of the paper is organized as follows. The simulator, mission, and experimental procedures are illustrated in Section~\ref{sec:design}. We briefly review evaluation methodology in Section~\ref{sec:method}. The finding of our experiments and hypotheses testing are provided in Section~\ref{sec:results}. We discuss our result and future work in Section~\ref{sec:conclusion}.

\section{Experiment Design and Procedure}\label{sec:design}

\subsection{Wind-aware Simulator-UI}
The wind-aware simulator is built upon three open-source software codebases; $1)$ AirSim~\cite{shah2018airsim}, $2)$ PX4~\cite{meier2015px4} and $3)$ QGroundControl (QGC) station~\footnote{http://qgroundcontrol.com/}. These platforms have been selected based on AirSim's capability to conduct software-in-the-loop (SITL) with PX4 and QGC. We use a PlayStation controller to integrate human pilot inputs into the system. While the original implementation of AirSim only allows global wind, we integrate an additional module to the core physics engine to handle local wind. Local wind changes based on the position of the quadcopter and simulation time. We also update the quadcopter dynamics to capture the effect of time-varying wind. The PX4 flight stack is modified to allow the transfer of local wind velocities to the QGroundControl station through the MAVLink communication protocol. PX4 communicates state observations to QGC for the pilot's interpretation. We use QML to develop wind widgets that display the
local wind velocity in a variety of ways. An example of wind widgets is provided in Fig 1. 
\begin{figure}[h!]
\centering
\begin{subfigure}{0.2\textwidth}
    \includegraphics[width=\textwidth]{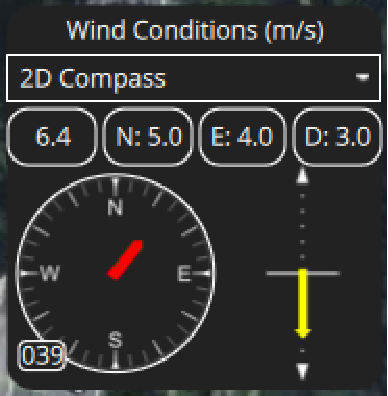}
        \caption{Compass-based heading (planer direction) with a vertical wind component.}
    \label{fig:2d_compass}
\end{subfigure}
\hfill
\begin{subfigure}{0.2\textwidth}
    \includegraphics[width=\textwidth]{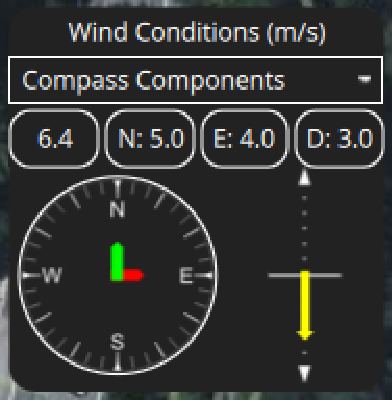}
          \caption{Component-wise display (north-east velocity split) with vertical wind component.}
    \label{fig:compass_comp}
\end{subfigure}
\hfill
\begin{subfigure}{0.2\textwidth}
    \includegraphics[width=\textwidth]{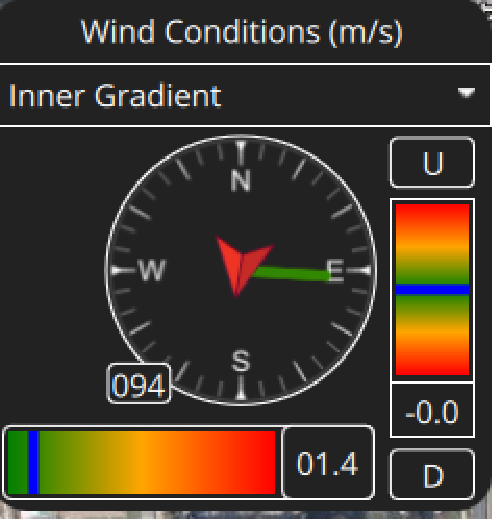}
         \caption{Gradients-based display with vehicle heading and wind direction in compass.}
    \label{fig:inner_grad}
\end{subfigure}
\hfill
\begin{subfigure}{0.2\textwidth}
    \includegraphics[width=\textwidth]{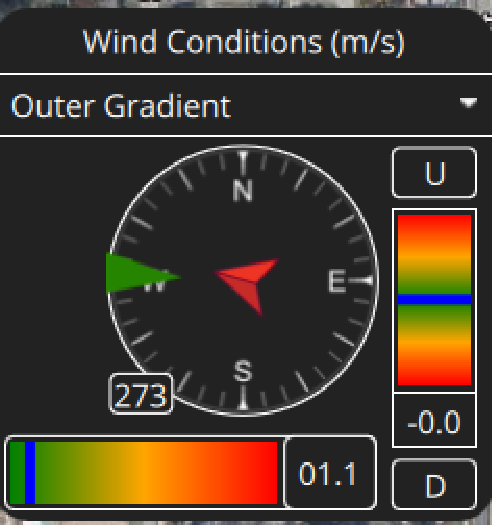}
        \caption{Gradients-based display with vehicle heading and wind direction in the outer dial.}
    \label{fig:outer_grad}
\end{subfigure}
\caption{Wind velocity module in wind-aware UI: pilot can choose preferred display from the drop-down menu.}
\label{fig:figures}
\end{figure}
Each interface conveys the velocity both numerically and graphically with the graphical implementation varying. Fig.~\ref{fig:2d_compass} uses a compass-style display, where the horizontal velocity of the wind is represented as an arrow conveying the wind's direction and magnitude. An alternative display is shown in Fig.~\ref{fig:compass_comp} with the planar velocity split into north and east vector components. Both displays show vertical wind with another arrow beside the compass. A third implementation is shown in Fig.~\ref{fig:inner_grad}, where the direction is conveyed using a compass, similar to Fig.~\ref{fig:2d_compass}. The magnitude is displayed on a green-yellow-red gradient beneath the compass. The final design refers to Fig.~\ref{fig:outer_grad}, similar to the third display with wind direction placed in the outer dial and indicates the relative direction of the wind with respect to the vehicle. 

\begin{figure}[h!]
    \centering
    \includegraphics[width=.7\textwidth]{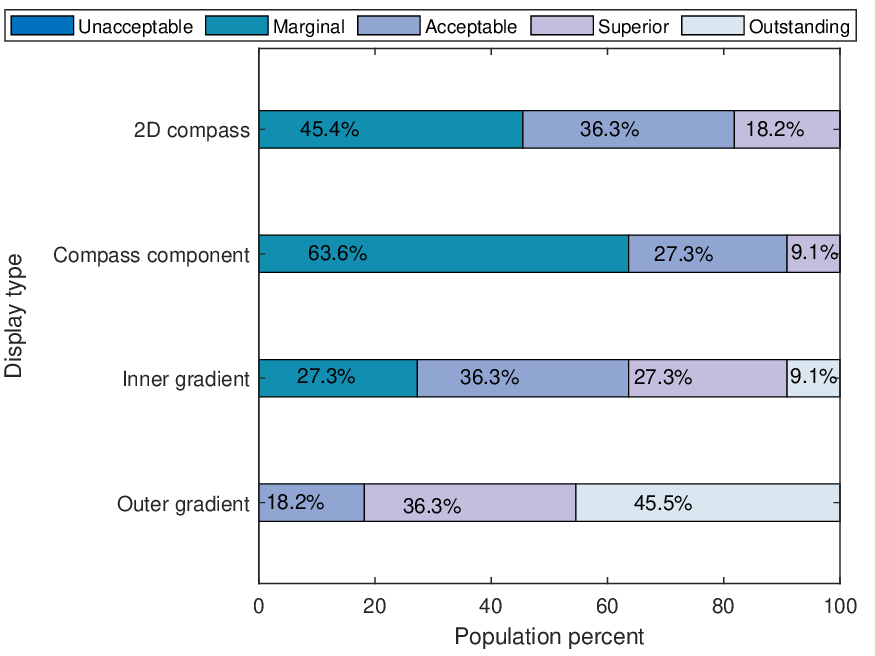}
    \caption{Likert plots showing survey on users' perspectives of different displays. Outer gradient where the wind can be interpreted relative to the vehicle position received the most positive vote. $9$ out of $11$ participants selected outer gradient for experiments.}
    \label{fig:likert}
\end{figure}

The displays are designed based on our SME's suggestions. During the experiment, we let the pilot choose their preferred display and let them scale the utility of each display on a $5$-point Likert scale~\cite{joshi2015likert} shown in Table~\ref{tab:likert}. A Likert scale allows an individual to express how much they agree or disagree with a particular statement. This enables us to gather feedback from the user which is a important part of user-centric design validation. We plot the voting scale of $11$ participants in terms of population percent in Fig.~\ref{fig:likert}. The outer gradient display that shows the wind direction with respect to vehicle heading received higher ratings than all other displays.
\begin{table}[h!]
    \centering
           \caption{Likert scale for feedback to wind display design. }
    \label{tab:logistic}
    \begin{tabular}{|p{0.2\textwidth}| p{0.7\textwidth}|}\hline
      Scale   & Meaning \\ \hline
      Unacceptable   & functionality is not sufficient to meet mission needs.\\
   Marginal & functionality sufficient to meet mission needs, but with at least one deficiency.
      \\
      Acceptable & functionality is sufficient to meet mission needs.\\
      Superior & functionality is sufficient to meet mission needs and exceeds at least one desired parameter. \\
       Outstanding & functionality exceeds all mission needs.\\
       \hline
    \end{tabular}
 \label{tab:likert}
\end{table}

\subsection{Mission Design}
\begin{figure}[h!]
    \centering
\includegraphics[width=.8\textwidth]{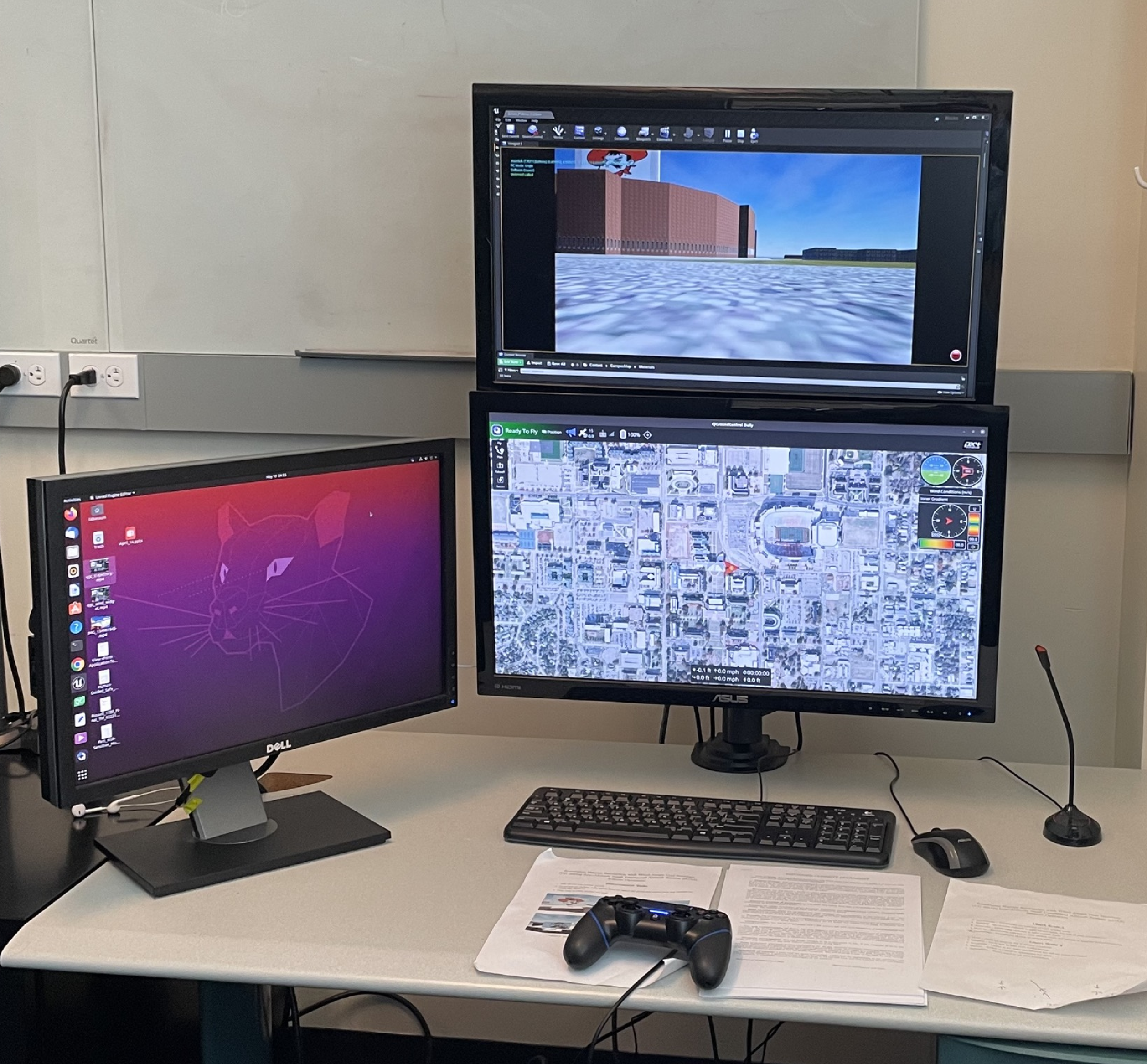}
    \caption{Experimental setup: upper screen shows the environment and the lower screen shows the ground control station. Participants are also required to use the mouse and the joystick.}
    \label{fig:exp setup}
\end{figure}

The experiment is carried out in the  simulator with the capacity of first-person view (FPV) through the camera.  An upper screen shows a video feed of the simulated scenario and a lower screen shows the wind-aware UI. This imitates the real flight setup scenario since the operator usually holds a screen with a controller in hand and looks up at the flight environment (head-down display). Fig.~\ref{fig:exp setup} shows the experimental setup with the controller. Experiments include conducting a mission in five different environment conditions and UI combinations. The experimental design matrix is provided in Table~\ref{tab:exp design}.
\begin{table}[h!]
    \centering
        \caption{Experimental design matrix.}
    \begin{tabular}{|c|c|c|} \hline
     Experiment & Wind condition  & display    \\ \hline 
      NW &  No wind  & No wind \\ 
      CW & Constant wind & No display \\ 

       TW &Turbulent wind & No display  \\ 
         CWD   &Constant wind & With display \\ 
    TWD & Turbulent wind & With display \\ \hline
    \end{tabular}
    \label{tab:exp design}
\end{table}
\subsubsection{Training}
 The participants are provided with a training session to help them understand the setup and tools available to them. This includes understanding and interpretation of wind displays as well. Different wind displays are explained from pictures and a video. The flight in simulations is fully manual and collision avoidance capability is not available. During training, participants also go through controller manipulation and are allowed as much as time they need to be comfortable with setups.

 \subsubsection{Main Experiment}
\begin{figure}[h]
     \centering
     \includegraphics[width=.75\textwidth]{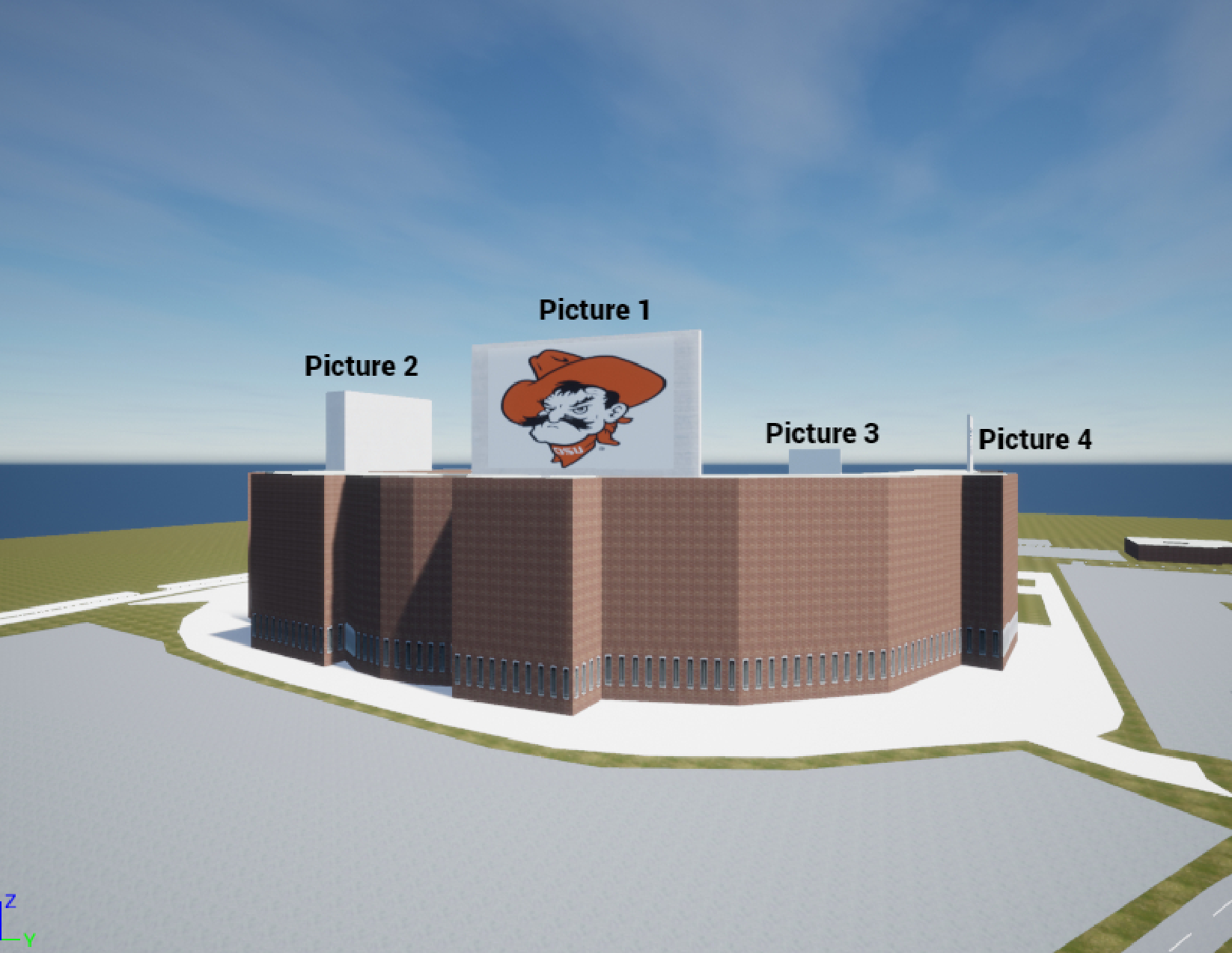}
     \caption{Mission: Go around the Boone Pickens Stadium and focus on the pictures at four different places.}
     \label{fig:mission 1}
 \end{figure}

 Experiments start after the participants feel confident about setups and control manipulation. The full set of experiments runs about 2 hours.  For each set of experiments, the operators are asked to achieve the following
Objectives in the simulations:
\begin{itemize}
    \item Take off and take manual control,
    \item Survey the stadium and focus on the assigned picture as steadily as possible. The whole environment is shown in Fig.~\ref{fig:mission 1},
    \item Come back to the home position and land.
\end{itemize}

\subsubsection{Participants}
\begin{figure}[h!]
      \centering
      \includegraphics[width=.8\linewidth]{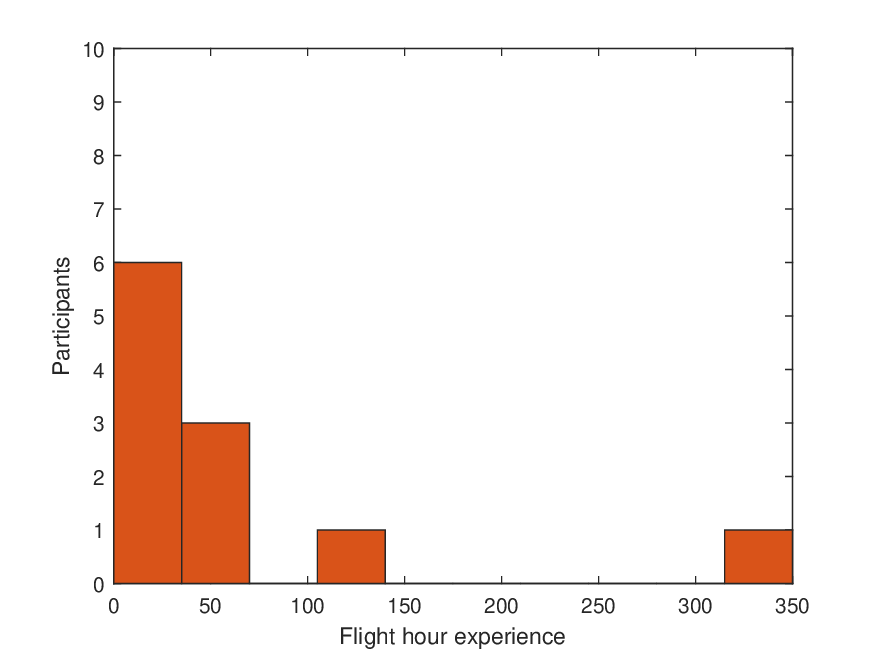}
       \caption{A histogram plot showing participants' flight experience. The mean and standard deviation of flight experience between participants are $M(11) = 61.54, SD(11) = 103.54$, respectively.}
    \label{fig:exp}
\end{figure}

As of June 18, a total of $11$ participants with varying flight experience, as shown in Fig.~\ref{fig:exp}, have participated in the experiment. Our recruitment and experiments are still ongoing.  The participants are recruited via email distribution list and snowball method with emails. The emails are circulated to the Mechanical and Aerospace Engineering, Electrical and Computer Engineering graduate students, and the College of Education and Human Sciences: Aviation students at Oklahoma State University. The experiment is approved by the Institutional Review Board (IRB) at Oklahoma State University. All participants were compensated with a  gift card. 
\subsubsection{Measurements}
Pre-experiment and logistic training surveys are conducted prior to starting the experiment. In the pre-experiment survey, we gather the flight experience data of the participants. The logistic survey questionnaire provided in Table~\ref{tab:logistic} is used to analyze  the operator's understanding of the setup and display. After the experiments, operators are given Subjective Work Assessment Technique (SWAT)~\cite{potter1989subjective} questionnaires for each set of experiments, a post-experiment survey, and a generic survey. The post-experiment survey consists of queries related to situational awareness assessment and wind information utility.  In the generic survey, the operator can provide feedback on the various features developed in this program and on how to improve them to better fit their needs in remotely piloting a UAS during windy conditions.  Along with the surveys from the participants, state and control input data are collected through the PX4 log file, and image data are automatically stored in  the computer. 
\begin{table}[h!]
    \centering
        \caption{Logistic questionnaires.}
    \label{tab:logistic}
    \begin{tabular}{|c|}\hline
      Questions    \\ \hline
       Are you able to interpret all the displays provided in the interface?  \\
       Which display serves best to understand the wind information? 
      \\
       Please identify the wind magnitude and direction of your chosen display \\
       \hline
    \end{tabular}

\end{table}

\section{Assessment Methods
}\label{sec:method}
\subsection{Cognitive Workload Assessment}
Workload assessment is a critical part of the study while designing a new interface or integrating layers of information into the interface. In our context, we evaluate mental workload since  our working environments impose more cognitive demands upon operators than physical demands. In highly dynamic environments with a multidimensional task load, an uncertain environment increases the operator's workload~\cite{hooey2017underpinnings}. In many situations, operators may have to increase and exhaust cognitive resources to maintain a high performance~\cite{reid1988subjective}. Several tools are used in the literature~\cite{young2015state} to measure and assess workload. Some tools are considered intrusive~\cite{reid1988subjective} that require operator interaction with designated tools. These tools may further increase the workload on top of the actual workload. Non-intrusive measurement tools have been widely used and accepted. In this measurement, operators are required to fill up the questionnaire at the end of the experiment sessions. Since the answers to the questionnaire are based on the operators' perceptions, these are called subjective assessments. Among several subjective assessment ratings, NASA-TLX~\cite{hart1988development}, Subjective Workload Assessment Technique (SWAT)~\cite{reid1988subjective}, and Bedford Scale~\cite{roscoe1990subjective} are frequently used in aerospace applications~\cite{jennings2005evaluating}. NASA-TLX and SWAT are both multi-dimensional methods: NASA-TLX comprises six subjective factors whereas SWAT comprises only three factors. Originally, SWAT has two phases. The first phase starts with sorting 27 cards with various combinations of workload, according to the individual perception of subjective cognitive loading in high to low order. A scale is then developed using the sorting order. The cognitive load is measured based on the questionnaire and developed scale. Bedford scale follows a flow pattern of questions, where answering to some question \textit{"yes"} prompts another question and \textit{"no"} ends the questionnaire. Although NASA-TLX and SWAT both incorporate a scaling procedure by obtaining the results from a number of individuals factor, we have chosen SWAT (see Table~\ref{tab:swat}) since it is based on a minimal scaling method while still producing meaningful results in relevant applications~\cite{zak2020subjective}. The modified SWAT is adopted where the pre-sorting and scaling steps are skipped and  the workload score is calculated based on the raw scores of the participants with Principal Component Analysis (PCA)~\cite{senders1979axiomatic}. 
\begin{table}[h!]
    \centering
        \caption{Subjective Workload Assessment Technique (SWAT) questionnaire.}
    \label{tab:swat}
    \small
    \begin{tabular}{|p{0.3\textwidth}| p{0.5\textwidth}|}\hline
       Load  & Scale and meaning \\ \hline 
    Time load  & \begin{tabular}[t]{ @{\makebox[1.5em][l]{\textbullet}} 
   p{\dimexpr\linewidth-1.5em} @{} }
    Often have spare time -  $1$ \\
       Occasionally have spare time - $2$   \\
        Almost never have spare time - $3$ \\
\end{tabular} \\
   Mental effort load & \begin{tabular}[t]{ @{\makebox[1.5em][l]{\textbullet}} 
   p{\dimexpr\linewidth-1.5em} @{} }
  Very little conscious mental effort or concentration required - $1$\\
        Moderate conscious mental effort required - $2$\\
        Extensive mental effort and concentration are necessary - $3$\\
\end{tabular}  \\
 Psychological stress load & \begin{tabular}[t]{ @{\makebox[1.5em][l]{\textbullet}} 
   p{\dimexpr\linewidth-1.5em} @{} }
        Little confusion, risk, frustration, or anxiety exists and can be easily accommodated - $1$\\
        Moderate stress due to confusion, frustration, or anxiety noticeably adds to the workload - $2$\\
       High to very intense stress due to confusion, frustration, or anxiety - $3$\\

\end{tabular}\\
 \hline
    \end{tabular}
\end{table}
\subsection{Situational Awareness Assessment}
Endsley~\cite{endsley1988design} defines situation awareness (SA) as being comprised of three components: $1)$ the perception of the elements in the environment within a volume of time and space (L1),~$2)$ the comprehension of their meaning (L2), and $3)$ the projection of their status in the near future (L3). SA is critical  for successful decision-making, especially in the aerospace domain~\cite{gibb2008classification}. Each of these levels may comprise specific situation awareness such as mission awareness, spatial awareness, and time awareness~\cite{gatsoulis2010measurement}. While effectively measuring SA poses considerable challenges due to its multivariate nature, it provides valuable information with higher subjective sensitivity. In the literature, three main methods had been utilized to measure and quantify SA. First, the explicit methods directly measure SA by assessing the features of participants' mental models concurrently with the tasks. One of the most sought methods is SAGAT (Situational Awareness and Global Assessment Technique)~\cite{stanton2004situation}, which is developed based on \textit{freeze technique}. In this assessment, the tasks are halted and assessment questionnaires are presented to the participants. While explicit methods are well accepted in the aviation domain, particularly in Air Traffic Control~\cite{endsley1994situation,kaber2006situation}, for safety-critical operations such as piloting airborne vehicles, obstructing or freezing tasks may not be an appropriate option. The second method is the implicit method that assesses situational awareness by inferring another intermediate variable. One common variable is task performance (TP)~\cite{neal1998human,muniz1998methodology}, which can be formulated based on mission accomplishments.  Predefined performance metrics may also be used. Despite being the minimal invasive assessment technique, this method may falsely tie poor performance with poor SA, as performance may vary for a variety of reasons, especially on the experience of the operator. The third method called the subjective method uses the participant's own judgment. The method is based on the principle that the participants know better. The assessment is conducted post-tasks and creates no obstruction during the experimental procedure. Due to the nature of self-assessment, this technique provides greater accuracy in understanding user-centric designs. Since we use an already established popular UI as our base display, we formulate SA questionnaires based on mission criteria. We follow {the third method and adapt} the Post Assessment of Situational Assessment (PASA)~\cite{gatsoulis2010measurement} structure while setting up the questions and  the 5-point rating scale and assigning the level of SA with each question (provided in Table~\ref{tab: sa}). 
\begin{table}[h!]
    \centering
        \caption{Post-experiment survey for subjective situational awareness assessment with response type and associated SA level.}
    \begin{tabular}{|p{0.60\textwidth} |p{0.15\textwidth}|p{0.12\textwidth} |p{0.08\textwidth} |}
\hline
       Questions & Basic/wind-aware & Answer/ rating & Level  \\
       \hline
       Were you able to know exactly where the quadcopter’s position was at all times? & Both & Rating & L1\\ 
       Were you able to identify building structures easily? & Both & Answer & L1 \\
       Were you able to keep track of time? & Both & Rating & L2 \\    
       Does display help you understand wind throughout the mission?& Wind-aware & Answer & L2
       \\ 
       How often do you use wind information to make decisions i.e. change control? Please rank for constant and turbulent wind separately. & Wind-aware & Rating & L3 \\
       Was it easy to follow the mission goal with the wind-aware display? Please rank for constant and turbulent wind separately. & Wind-aware & Rating & L3 \\
       Were you able to change the course of action because you felt more confident with the wind-aware display than with a basic display?& Wind-aware & Rating & L3 \\
       \hline
    \end{tabular}
    \label{tab: sa}
\end{table}
\section{Result Analysis}\label{sec:results}
\begin{figure}[h!]
\centering
\begin{subfigure}{0.48\textwidth}
\includegraphics[width=\textwidth]{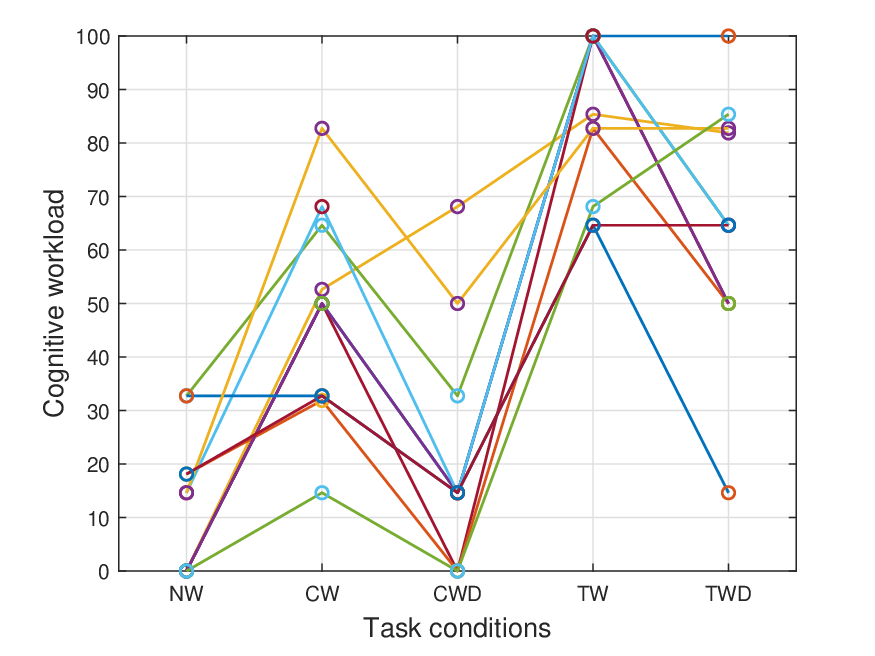}
     \caption{Cognitive workload of participants $(N=11)$. }
 \label{fig:swat_all}
\end{subfigure}
\hfill
\begin{subfigure}{0.48\textwidth}
    \includegraphics[width=\textwidth]{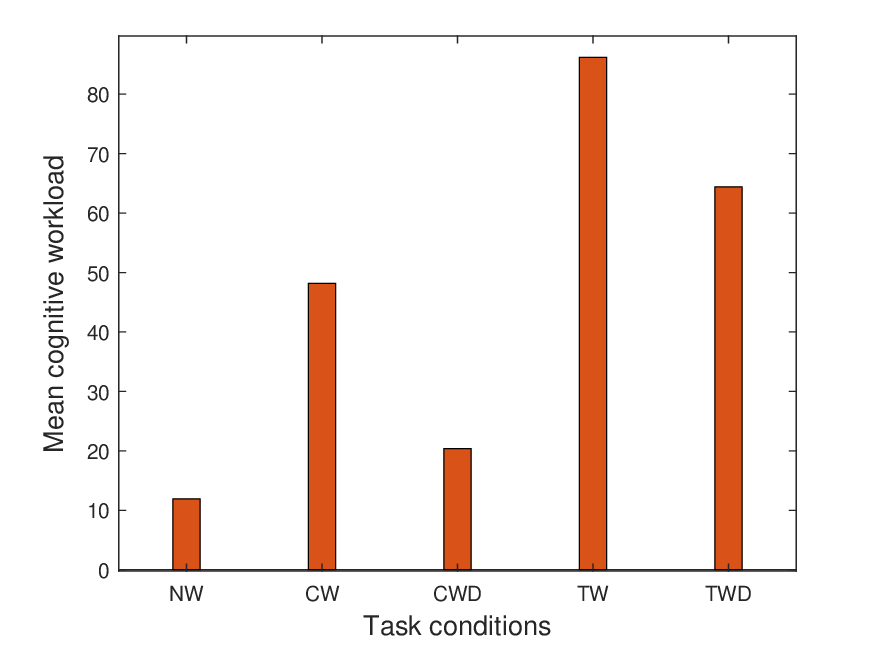}
    \caption{Mean cognitive workload for each task condition.}
  \label{fig:mean_swat}
\end{subfigure}
\hfill
\caption{Cognitive workload: (a) shows individual workload and (b) shows mean workload for each task condition. Both figures indicate that with display the  cognitive workload tends to decrease in both constant and turbulent task conditions.}
\label{fig:figures}
\end{figure}

To study the usability of the wind-aware display we formulate the following null hypotheses.
\begin{hyp}[$H1_{0}$] \label{hyp:one}
There is no significant difference between cognitive load in different wind conditions.
\end{hyp}
\begin{hyp}[$H2_{0}$] \label{hyp: two}
There is no significant difference between cognitive load in different display conditions.
\end{hyp}
\begin{hyp}[$H3_{0}$] \label{hyp: three}
There is no significant difference between situational awareness in different display conditions.
\end{hyp}
To assess the hypothesis we adopt a one-way ANOVA analysis with $95\%$ confidence interval setting. The cognitive workload is calculated from the raw scores provided by the participants. We skip the time-consuming sorting card stage and utilize PCA instead. The PCA scores are used to deduce the total cognitive load which is then normalized to achieve a workload value spanning from $0$ to $100$. The overall cognitive workload is illustrated in Figure~\ref{fig:swat_all} and the mean workload is shown in~\ref{fig:mean_swat}.
\begin{figure}[h!]
\centering
\begin{subfigure}{0.48\textwidth}
\includegraphics[width=\textwidth]{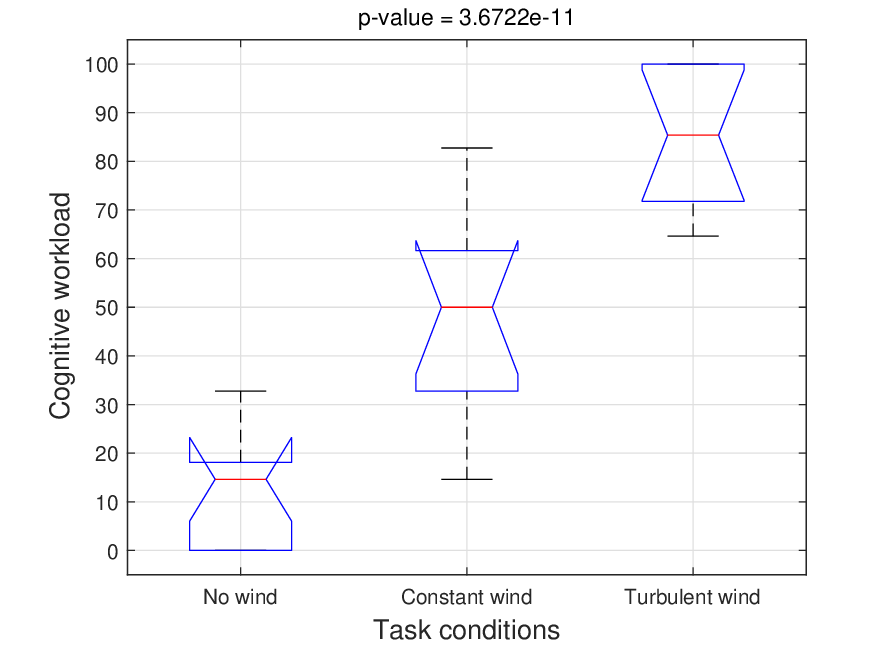}
     \caption{Cognitive workload in the different wind conditions without display.}
    \label{fig:wind_swat}
\end{subfigure}
\hfill
\begin{subfigure}{0.48\textwidth}
    \includegraphics[width=\textwidth]{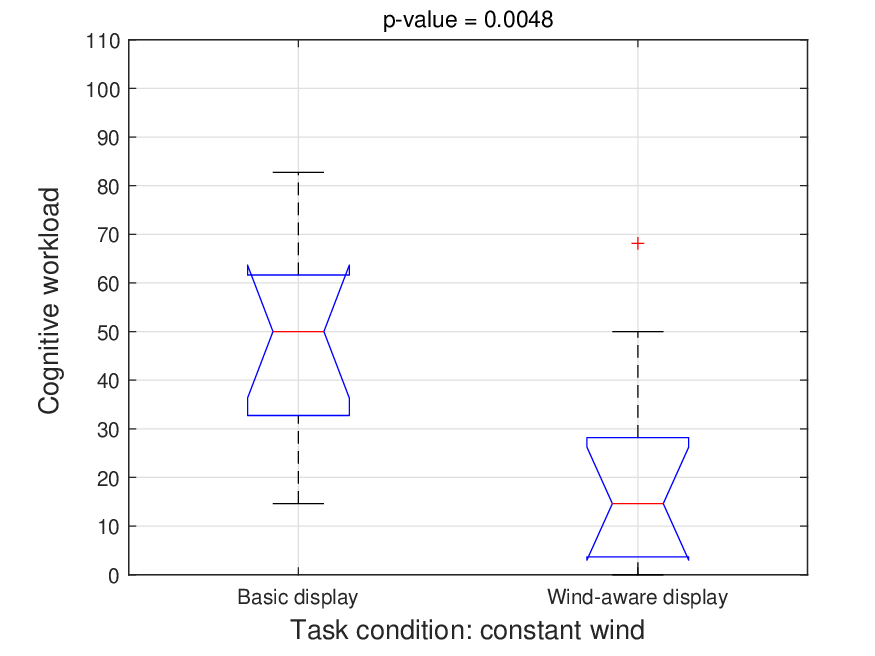}
      \caption{Cognitive workload in the constant wind with different display conditions.}   \label{fig:con_swat}
\end{subfigure}
\hfill
\begin{subfigure}{0.48\textwidth}
    \includegraphics[width=\textwidth]{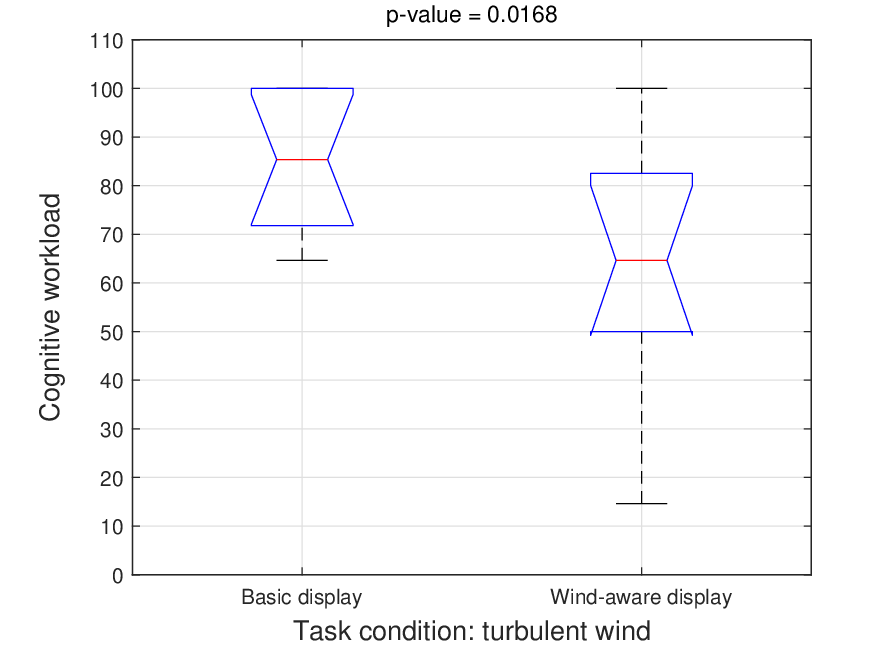}
    \caption{Cognitive workload in the turbulent wind with different display conditions.}
    \label{fig:turb_swat}
\end{subfigure}     
\caption{Illustration of ANOVA results in box-plot. The red line represents the median of the data, and the starting, and ending of the notch represent the first and third quartiles of the data. Here, (a) clearly outlines the variability of the workload between wind conditions. In plots (b) and (c) we also distinctly notice the difference in workload data spans.}
\label{fig:figures}
\end{figure}
 Both plots imply the fact that the complexity of wind increases participants' cognitive workload. With ANOVA analysis, we see a statistically significant difference in workload $(p-value = 3.67 \times 10^{-11} < 0.05)$ in different wind conditions (see Figure~\ref{fig:wind_swat}). A summary of the cognitive workload analysis and hypotheses results are provided in Table~\ref{tab:anova workload}. The F-statistics for hypothesis~\ref{hyp:one}, $F(2,11) = 59.43$ signify how large the variance between workloads are in the different wind.  We also observe that providing the participants with the wind information alleviates the cognitive workload significantly  $(p-value < 0.05)$ in both constant and turbulent wind conditions (refer to figures~\ref{fig:con_swat} and \ref{fig:turb_swat}).  ($H2_{0}$). 
\begin{table}[h!]
    \centering
        \caption{Descriptive statistics of ANOVA analysis of cognitive workload for $N=11$ participants.}
    \label{tab:anova workload}
    \begin{tabular}{|c|c|c|c|}
    \hline
       Hypothesis  &  F-value & p-value & Decision \\
       \hline
      $H1_{0}$: No wind  & $59.43$ & $3.67\times10^{-11}$ & Reject\\
      $H2_{0}$: Constant wind & $10.05$ & $0.0048$ & Reject \\
       $H2_{0}$: Turbulent wind & $6.8$ & $0.00168$ & Reject \\
       \hline
       \end{tabular}
\end{table}
\begin{figure}
    \centering
    \includegraphics[width=.6\textwidth]{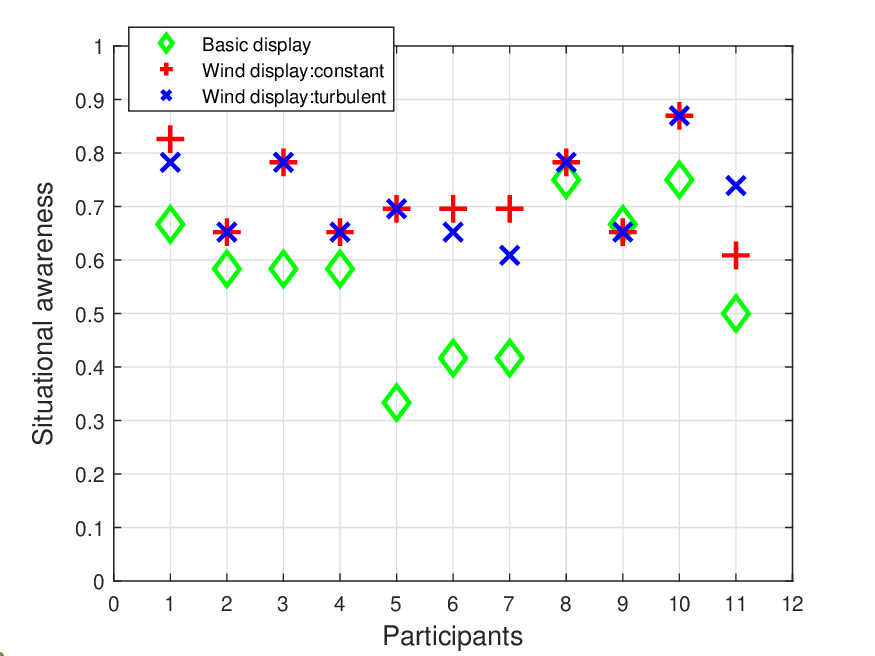}
    \caption{Subjective situational awareness score for each participant in different wind conditions. Some participants feel wind-aware display improved SA equally during constant and turbulent wind conditions which is depicted by overlapping red and blue points. }
    \label{fig:individual_sa}
\end{figure}
\begin{figure}[h!]
\centering
\begin{subfigure}{0.49\textwidth}
    \includegraphics[width=\textwidth]{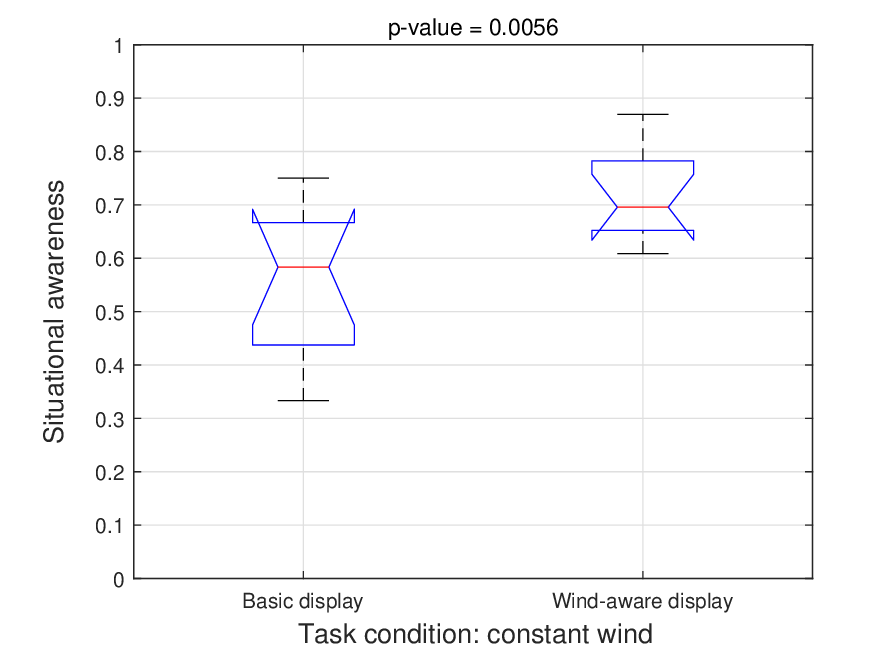}
     \caption{Subjective situational awareness in the constant wind with different display conditions.}   \label{fig:con_sa}
\end{subfigure}
\hfill
\begin{subfigure}{0.49\textwidth}
    \includegraphics[width=\textwidth]{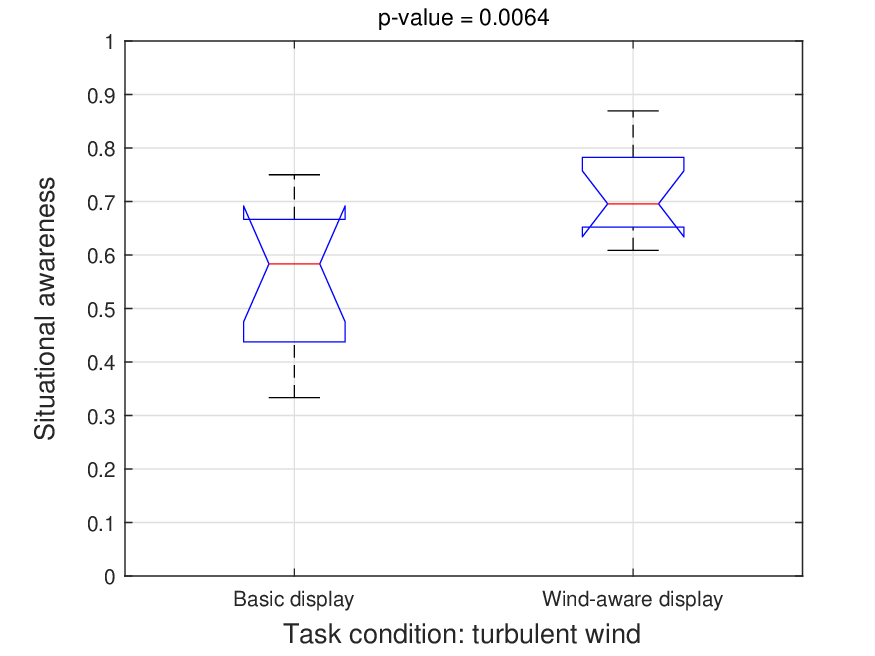}
       \caption{Subjective situational awareness in the turbulent wind with different display conditions.}
    \label{fig:turb_sa}
\end{subfigure}
\hfill
\caption{Illustration of ANOVA results in box-plot. It should be noted that the data spans for constant and turbulent wind look similar from the box plots (a) and (b), as most of the participants rated SA the same for both constant and turbulent wind. }
\label{fig:sa}
\end{figure}

We next evaluate the null hypothesis~\ref{hyp: three} ($H3_{0}$) for situational awareness. To do that, we first add up the answer and ratings provided by the participants post-experiments. The questionnaires are categorized into both basic and wind-aware displays (see Table~\ref{tab: sa}). The score is normalized on a scale of 1. The subjective situational awareness scores for each participant is shown in Fig.~\ref{fig:individual_sa}. While most participants feel the acquired SA is the same for constant and turbulent wind, a few participants rated SA differently for constant and turbulent wind conditions, which is visible in Fig.~\ref{fig:individual_sa}. 
Table~\ref{tab:anova sa} shows the ANOVA statistics for the post-assessment of situational awareness questionnaires, where we include the $F$ and $p$ values. 
\begin{table}[h!]
    \centering
      \caption{Descriptive statistics of ANOVA analysis of subjective situational awareness for $N=11$ participants.}
    \label{tab:anova sa}
    \begin{tabular}{|c|c|c|c|}
    \hline
   Hypothesis  & F value & p-value & Decision  \\
   \hline
   $H3_{0}$: Constant wind   & $9.62$ & 0.0056 & Reject\\
    $H3_{0}$: Turbulent wind &  $9.26$ &  0.0064 & Reject\\
    \hline
\end{tabular}
\end{table}
The ANOVA analysis shows that wind information significantly improves subjective situational awareness in both constant $(p-value = 0.0056 < 0.05)$  and turbulent wind situations $(p-value = 0.0064 < 0.05)$. The box plots in Fig.~\ref{fig:sa} visualize the impacts of the wind information in improving situational awareness. The $F$ values are very close for constant wind $(F(1,11) = 9.62)$ and turbulent wind  $(F(1,11) = 9.26)$ with the constant wind case being slightly higher. This is because the participants who reported different SA for constant and turbulent wind indicated higher SA acquired during constant wind operations except for participant $P11$. Some of the SA questionnaires are based on if participants were able to utilize wind information for decision-making. A few participants expressed that due to the higher cognitive loading on quadcopter control during turbulent operations, oftentimes they were unable to make use of wind information provided at a desired level, which could be the reason that we see a lower SA acquired in the turbulent case than in the constant case.

\section{Discussion and Future Work}\label{sec:conclusion}
The preliminary post-hoc analysis exclusively focuses on understanding subjective perspectives. Although the sample size is small, consistent outcomes are noticed while analyzing the results in varying sample sizes starting from $N=6$. Hence, we expect a similar outcome as we continue experiments with more participants. Participants were also asked what they liked or disliked about the experiment. The most positive remark is that the wind display makes them more confident. It is also noted that the pilots benefited more from the direction than the magnitude. In real time while there is turbulence during taking pictures, most of the pilots use wind direction while projecting the outcome of their decision, not the magnitude. All participants mention they would take assistance from autonomous control if available. One of the participants, who frequently operates some sUAS missions mentions that the first-person view helped carry out the mission along with the wind display. 

The current study provides subjective or user insights using a wind-aware display.  It is also important to analyze the objective or mission-specific improvement with the wind-aware display. Future work includes analysis of flight and image data to understand objective improvement and measure the  performance of the pilot under different conditions. 
\printbibliography

%
%

%
%
%
%

\end{document}